\documentstyle[12pt]{article}
\begin{document}

\begin{center}
{\Large\bf
QED radiative corrections to impact factors
\vskip 5mm
E.A. Kuraev, L.N. Lipatov, T.V.Shishkina
}
\end{center}

\begin{abstract}
We consider the radiative corrections  to the impact factors of
electron and photon.
According to a generalized eikonal representation the
$e^+e^-$ scattering amplitude at
high energies and fixed momentum transfers   is proportional
to the electron formfactor. But we show that this representation is
violated due to the presence of non-planar diagrams.
One loop correction to the photon impact factor
for small virtualities of the exchanged photon
is obtained using the  known results for  the
cross section of the $e^+e^-$ production at photon - nuclei
interactions.
\end{abstract}

\section{Introduction}
It is well known (see \cite{CWGLF}), that for the QED scattering
amplitude of the process $a+b \rightarrow a'+b'$ in the
Regge kinematics
\begin{eqnarray}
A(p_A,a)+B(p_B,b)\to A(p_A',a')+B(p_B',b'), \\ \nonumber
s=(p_A+p_B)^2 >>-t=-(p_A-p_A')^2\sim m^2,
\end{eqnarray}
the simple representation
\begin{eqnarray}
A(s,t)=\frac{i s}{(2\pi)^2}\int\frac{d^2k \, \,\tau^A(k,r)\, \tau^B(k,r)}
{[(k+r)^2+\lambda^2][(k-r)^2+\lambda^2]}(1+O(\frac{t}{s})),\,\,
4r^2=-t<0\,,
\end{eqnarray}
is valid in the first non-trivial order of the perturbation theory.
Here $\lambda$ is the photon mass and the impact factors (IF) $\tau$
describe the inner
structure of
colliding
particles. The quantities $r$ and
$k$ are
two-dimentional vectors orthogonal
to the momenta  $p_A, p_B$ of initial particles.
For the electron $|\tau^e|=4\pi\alpha \delta_{ij}$,
where
indices $i,j$ describe its polarization states. The
expression for IF of the photon on its mass shell can be presented in
the form~\cite{CWGLF}:
\begin{eqnarray}
\tau^{\gamma}_{ij}=8\alpha^2\int_0^1 d y\int_0^1 dx_+ dx_-
\delta(x_++x_--1)(A_{ij}-B_{ij}),
\end{eqnarray}
with
\begin{eqnarray*}
A_{ij}=\frac{1}{4r^2x_+^2 y(1-y)+m^2}\!\!\!\!&&\!\!\!\![8x_+^3x_-y(1-y)r_ir_j
\\
&-& x_+^2r^2(1-8x_+x_-(y-\frac{1}{2})^2)\delta_{ij}]; \\ \nonumber
\end{eqnarray*}
\begin{eqnarray*}
B_{ij}&=&\frac{1}{4Q^2 y(1-y)+m^2}[8x_+x_-y(1-y)Q_i Q_j-
Q^2(1-8x_+x_-(y-\frac{1}{2})^2)\delta_{ij}], \\ \nonumber
Q&=&\frac{1}{2}(k+r)-x_+ r \,,
\end{eqnarray*}
where  $i,j$ descibe the photon polarization states.

{}From the point of view of the Regge theory the impact factors are
proportional to the residue at $j=1$ of the pole of the $t$-channel
partial wave $f_j$
describing the transition of two particles into  the nonsence state
of two virtual photons  \cite{CWGLF}. In the upper orders of the
perturbation theory there appear the poles $f_j\sim 1/(j-1)^n$
which should be subtracted from $\tau$ providing one sums all
logarithmic cotributions $\sim \log^n (s)$ with the use of the
Bethe-Salpeter equation  \cite{CWGLF}.

For $t=0$ the impact factor is proportional to the integral from
the total cross-section for the scattering of the virtual photon
off the target. In the particular case, when the photon transverse
momentum $k$ tends to zero, the impact factors can be expressed in
terms of the integrals from the total cross-section for the real
photon
\begin{eqnarray}
\tau = k^2 \,\int _{th}^{\infty} \frac{d \,s}{\pi} \frac{\sigma
_{a\gamma }(s)}{s}\,,
\end{eqnarray}
which corresponds to the Weizsaekker-Williams approximation.

The motivation for our calculation of radiative corrections(RC) to
impact factors is the high presision experiments performed
on colliders where some physical quantities (for example,  the BFKL
Pomeron intercept) are measured  \cite{BFKL}. In this case one
should know IF of
the virtual photon  \cite{BFKL}.  Generally IF describe the coupling
of the
particles with the Pomeron in QED or in QCD.
For colliders with electron (positron) beams our results can be used
to calculate the QED-part of cross-sections with a good accuracy.
It turns out that the scattering amplitude corresponding to the diagrams
with the two
photon exchange is purely imaginary. In the case of the small
angle $e^+e^-$
scattering  the amplitude for the diagrams with the multi-photon
exchange has the eikonal representation
\begin{eqnarray}
&&A(s,t)=A_0(s,t)e^{i\delta(t)}, \,\, A_0(s,t)=4\pi\alpha\frac{2}{st}
\bar{u}(p_1')\hat {p}_2u(p_1)\bar {v}(p_2)\hat{p}_1v(p_2')=
4\pi\alpha\frac{2s}{t}N_1N_2,  \nonumber  \\
&&|N_i|=1,\delta(t)=-i\alpha\ln\frac{-t}{\lambda^2}\,.
\end{eqnarray}
Here we used the fact, that only longitudinal
(nonsence) polarizations of the
$t$-channel virtual photons are essential
\begin{eqnarray}
\bar{u}(p_1')\gamma_\mu u(p_1)\bar {v}(p_2)\gamma_\nu v(p_2')G^{\mu\nu}(q),\,\,
G^{\mu\nu}(q)=\frac{1}{q^2}\,\frac{2\,p_2^\mu p_1^\nu}{s}\,.
\end{eqnarray}
RC to $A_0$ appear from the so-called "decorated boxes"
(see Fig.1). These Feynman diagrams were assumed to lead to a
generalized
eikonal representation
\begin{eqnarray}
 A=A_0(s,t)[\Gamma_1(t)]^2e^{i\delta(t)},
\end{eqnarray}
where $\Gamma_1(t)$ is the Dirac form factor of electron
\begin{eqnarray}
V^\mu(t)=\gamma^\mu\Gamma_1(t)+\frac{\sigma^{\mu\nu}q_\nu}{2m}\Gamma_2(t),\, \,
q^2=t,\nonumber \\
\Gamma_1(t)=1+\gamma \Gamma_1^{(2)}(t)+...,\,\, \gamma=\frac{\alpha}{\pi}.
\end{eqnarray}
Note, that one should include in $\delta (t)$ also corrections to
the virtual photon Green function, leading in particular to the
electric charge renormalization.

To begin with, in the next section we verify the generalized eikonal
representation for the decorated boxes.

\section{RC to IF of electron}

Keeping in mind that the amplitude for the near forward scattering with
the two photon exchange is pure imaginary
(we omit the corrections of order $m^2/s$) we
can calculate its s-channel discontinuity. There are RC
to this discontinuity from the virtual photons and  from  the 
emission of the real photon in the
intermediate state. The last contribution we will
separate in two parts corresponding to  the emission of soft and
hard photons.

The virtual photon contribution contains the
electron vertex function  for the case when the initial and final
electrons are on mass shell:
\begin{eqnarray}
\Delta\tau_e^{virt}&=&\frac{\alpha}{\pi}\tau_e^(0)[F_1^{(2)}(k^2) +
F_1^{(2)}(k^{'2})],  \nonumber \\
F_1^{(2)}(t)&=&\ln\frac{m}{\lambda}
(1-\frac{1+a^2}{2a}\ln b) - 1 + \frac{1+2a^2}{4a}\ln b   \nonumber   \\
&-& \frac{1+a^2}{2a}
[-\frac{1}{4}\ln^2 b + \ln b\ln(1+b) - \int_1^b\frac{dx}{x}\ln(1+x)],
\nonumber \\
a&=&\sqrt{1-\frac{4m^2}{t}}, \qquad b=\frac{a+1}{a-1}, \,t<0.
\end{eqnarray}

The contribution from the emission of a soft photon have the
classical form:
\begin{eqnarray}
-\frac{\alpha}{4\pi^2}(\frac{p_1}{p_1k_1}-\frac{p}{pk_1})
(\frac{p_1}{p_1k_1}-\frac{p_1'}{p_1'k_1})\tau_e^{(0)}
\times \frac{d^3k_1}{\omega_1}|_{\omega_1<\delta E},\,\,\delta E<<E=\sqrt{s}/2,
\end{eqnarray}
where the momenta of initial and final electrons are
$p, p_1'$ and the
momentum of electron in the intermediate state is $p_1$. Because
the energies of these particles are approximatelly equal (and large
in comparison with the electron mass) we can use the relations:
\begin{eqnarray}
\frac{1}{2\pi}\int\frac{d^3k_1}{\omega_1}\frac{m^2}{(p_ik_1)^2}&=&2L_e;
\nonumber \\
\frac{1}{2\pi}\int\frac{d^3k_1}{\omega_1}\frac{p_1p_2}{(p_1k_1)(p_2k_1)}&=&
\frac{1+a^2}{a}[L_e\ln b -\frac{1}{4}\ln^2 b  \nonumber \\
&+&\ln b\ln(1+b)-\int_1^b\frac{dx}{x}\ln(1+x)], \nonumber \\
L_e&=&\ln\Delta+\ln\frac{m}{\lambda}, \,\, t=(p_1-p_2)^2, \
\Delta=
\frac{\delta E}{E} \ll 1,
\end{eqnarray}
with the quantities $a,b$ defined in (9).

At last we consider  the hard photon emission. Its
contribution to the
imaginary part of the electron-electron scattering amplitude
can be presented in the form
\begin{eqnarray}
Im_sA(s,t) = -s\frac{\alpha^3}{2\pi^2}\int\frac{d^2k}{k^2k^{'2}}N_1N_2
\frac{d^2k_1 dx}{x(1-x)}I(x,k_1,k), \ \Delta < x < 1,
\end{eqnarray}
where $x$ is the  energy fraction of the hard photon. We obtain
\begin{eqnarray}
I(x,k_1,k)&=&\frac{1}{d_1d_2}(-4m^2+2t_1z)
+ \frac{1}{d_1d_1'}(-4m^2x^2(1-x)+2tz(1-x)) \\ \nonumber
&+&
\frac{1}{d_2d_1'}(-4m^2+2t_2z)-2z\frac{1}{d_1}-4z\frac{1}{d_2}
+\frac{8m^2}{d_2^2}
-2z\frac{1}{d_1'}, z=1+(1-x)^2,
\end{eqnarray}
where
\begin{eqnarray}
d_1&=&(P-k_1)^2-m^2=-\frac{1}{x}[m^2x^2+\vec{k}_1^2],  \\ \nonumber
d_2&=&(p_1+k_1)^2-m^2=\frac{1}{x(1-x)}[m^2x^2+(x\vec{k}-\vec{k}_1)^2],\\
\nonumber
d_1'&=&(P'-k_1)^2-m^2=-\frac{1}{x}[m^2x^2+(x\vec{q}-\vec{k}_1)^2].
\end{eqnarray}
The subsequent  integration is straightforward and gives the result:
\begin{eqnarray}
\Delta\tau_e^{hard}&=&\tau_e^{(0)}\frac{\alpha}{\pi}[\ln\frac{1}{\Delta}(G(k^2)
+
G(k^{'2})-G(t))+G_1(k^2)+G_1(k^{'2})-G_1(t)], \nonumber \\
G(t)&=&\frac{1+a^2}{2a}\ln b -1;G_1(t)=1-\frac{1+2a^2}{4a}\ln b.
\end{eqnarray}
The interference of two amplitudes with the photon emitted by two
initial particles is small  $\sim O(t/s)$.
This fact is known in literature as the up-down cancellation.
The contribution of the diagrams with the two photon exchange is
pure
imaginary and, consequently,
does not interfere with the
real  Born amplitude.
Adding together all contributions we obtain the final result for one
loop RC to the
electron IF
\begin{eqnarray}
\Delta\tau_e=-\frac{\alpha}{\pi}\tau_e^{(0)}F_1^{(2)}(t),\,\,\tau_e^{(0)}=
4\pi\alpha.
\end{eqnarray}
This result agrees with the generalized
eikonal form of the small angle scattering amplitude. But in the upper
orders the eikonal representation is violated, as it will be shown below.

\section{Discussion}

The above result for RC to the electron IF can be obtained in a simple
way.
For simplicity let us consider the case when the positron block
consists only from the single fermion line whereas  electron one
contains the
set of 4 Feynman graphs, describing the decorated box.
We express the components of the exchanged photon momentum in terms of
the squared invariant energies $s_1,s_2$ for electron and positron blocks
\begin{eqnarray}
k=\alpha p_2+\beta p_1+k_\bot; \,\,
d^4 k=\frac{s}{2}d\alpha d\beta d^2 k_\bot=\frac{1}{2s} d s_1 d s_2
d^2 k_\bot; \nonumber   \\
s_1=(k-p_1)^2=s\alpha-\vec {k}^2; \,\, s_2=(k+p_2)^2=s\beta-\vec {k}^2,
\nonumber
\end{eqnarray}
where the Sudakov parametrization is used.
Performing the $s_2-$integration by  residue from the propagator of the
intermediate positron (it takes into account also the diagram with the
crossed photon lines), we obtain the following expression for the total RC
\begin{eqnarray}
-\frac{8i\alpha^2}{st}\int\frac{d^2\vec{k}\ \vec{q}^2}{(\vec{k}^2+\lambda^2)
((\vec{q}-\vec{k})^2+\lambda^2)}\bar{v}(p_2)\hat{p}_1v(p_2^{'})\int_C d s_1
p_2^\mu p_2^\nu\bar{u}(p_1')A_{\mu\nu}u(p_1),
\end{eqnarray}
where $\bar{u}(p_1')A_{\mu\nu}u(p_1)$ is the Compton scattering amplitude,
corresponding to the Feynman diagrams having only $s$-channel
singularities
and the contour $C$ is situated above these singularities. The
amplitude has the pole at $s_1=m^2$ which corresponds to the electron
intermediate state and the right hand cut starting from
$s_1=(m+\lambda)^2$, which
corresponds to one electron and one photon intermediate state.

Using the Sudakov
representation for the photon momentum $k$ we can present $p_2^\mu$ in
the
form
\begin{eqnarray}
p_2^\mu=\frac{1}{\alpha}(k-k_\bot-\beta p_1)\approx
-\frac{s}{s_1+\vec{k}^2}(k-k_\bot)^\mu.
\end{eqnarray}
Let us consider the product of two terms in right side part of
eq.(17) with the Compton amplitude.
The contribution of the first term$\sim k_\bot$ is zero:
\begin{eqnarray}
|\vec{k}|s p_2^\nu\int_C\frac{d s_1 k_\bot^\mu}{(s_1+\vec{k}^2)|\vec{k}|}
\bar{u}(p_1')A_{\mu\nu}(s_1,k,k')u(p_1)=0.
\end{eqnarray}

This conclusion follows from  the convergence of the integral
over the large circle in the $s_1$ plane and the absence of the left
cut.
The second property
is valid for the planar Feynman graphs. The convergence of the
integral follows from the fact that only 
physical transverse polarizations of the
virtual photon with
the momentum $k$ contribute and  the quantity $e^\mu
p_2^\nu
A_{\mu\nu},\vec{e}=
\vec{k}_\bot/|\vec{k}|$ behaves at large $s_1$ as $m^2/s_1$.

The cancellation of contributions from the pole  and the cut
for
planar graphs is the base for sum rules which relate cross
sections
of various production amplitudes in the Weizsaekker-Williams approximation
and the slope of
Dirac formfactor at zero momentum transfer(see \cite{BFKK}).

Applying the Ward identity to the first term $k$ in (17)  we
obtain:
\begin{eqnarray}
p_2^\mu p_2^\nu \bar{u}(p_1')A_{\mu\nu}(s_1)u(p_1)=-\frac{se^2}{s_1}
p_2^\mu\bar{u}(p_1')\Gamma^{\mu}(q)u(p_1), \,\,s_1>>m^2.
\end{eqnarray}
Now the large circle integral gives the generalized eikonal result,
which means in particular, that for arbitrary $t$
the total contribution of the various intermediate states is not
zero.

We see that  RC to IF of electron contain infrared divergences, which
are cancelled in the cross-section  with the contribution of
the inelastic process --
the photon emission.

RC to the photon IF do not contain infrared singularities, but unfortunately
we do not know the closed expression for them. We can estimate only
their size  at small virtualities of exchanged photon $\vec{k}^2$
and $q=0$. This value can be extracted from the
results of paper \cite{VKM}, where the one-loop correction to the cross
section of
pair production by photon on the coulomb field of nuclei was calculated:
\begin{eqnarray}
\sum_{i=1}^{i=2}[\tau+\Delta\tau]_{ii}^\gamma(k,0)=
\frac{28\vec{k}^2\alpha^2}{9m^2}[1+\delta_p], \,\,
\vec{k}^2<<m^2, \nonumber \\
\delta_p=\frac{\alpha}{\pi}\frac{9}{14}(\frac{1128}{35}\zeta(3)-
\frac{6971}{210})=0.009.
\end{eqnarray}

The contribution of the Pauli form factor in the lowest order of PT
is
suppressed by the factor $O(t/s)$, but it survives in higher orders
of PT.
Due to the known property of Dirac form factor $F_1(0)=0$ we obtain (in
an agreement
with the results of paper\cite{KLMF}) the vanishing of RC for the
case $t=0$.


The generalized eikonal (GE) hypothesis is violated in the 2-loop
approximation to IF \cite{KLM}).
This fact can  be verified for the case $t = 0$. If the GE
hypothesis would work, the complete compensation of contributions
to the electron IF from different processes would take place. We
show now that such compensation is absent on the
2-loop level. Let us onsider the high-energy small-angle
electron-positron
scattering amplitude corresponding to the "decorated"
three-loop diagrams
with the two-photon
exchange in the scattering channel. It has the imaginary part only in the
$s$-channel  (the diagrams with crossed photons  in
t-channel give the same contribution). We present the amplitude
in the form
\begin{eqnarray}
A(s,t) = \int\frac{d^{4}k}{(2\pi)^4} \frac{J^{(A)}_{\mu\nu}
J^{(B)}_{\mu_{1}\nu_{1}}}{k^{2}(q-k)^{2}}g^{\mu\mu_{1}}
g^{\nu\nu_{1}}.
\end{eqnarray}
Using the above simplification for the Green functions of
the exchanged photons
$$
g^{\mu{\mu}_{1}}\approx \frac{2}{s}p_A^{\mu_1}p_B^{\mu};\,\,\,
g^{\nu{\nu}_{1}}\approx \frac{2}{s}p_A^{\nu_1}p_B^{\nu},
$$
and the Sudakov parametrization of the exchanged photon momentum
in terms of the squared invariant masses
$s_1=s\alpha$ and $s_2=-s\beta$
of the particles moving
along directions of the initial particles $p_A$ and $p_B$ respectivelly,
we arrive
to the impact picture of the scattering amplitude (1) with:
\begin{eqnarray}
\tau^A=
\int_C\frac{d s_1}{2\pi}\frac{1}{s^2}J^{(A)}_{\mu\nu}p_B^\mu p_B^\nu;\,\,\,
\tau^B=
\int_C\frac{d s_2}{2\pi}\frac{1}{s^2}J^{(B)}_{\mu\nu}p_A^\mu p_A^\nu
\,,
\end{eqnarray}
where the
quantities $(1/s^2)J^{(i)}_{\mu\nu}p_j^\mu p_j^\nu,$  do not
depend on $s$ in the limit of large $s$. The integration contour $C$
is displaced in correspondence with the Feynman prescription between
the right and left hand side singularities of the amplitude.

As above $\tau^{A}$ has a pole  corresponding to  a
single particle  intermediate state,
the right and left cuts, corresponding to the intermediate states
with 2 and 3 particles.

For planar diagrams the
left cut in $s_1$-plane is absent. First time  it appears at the
two-loop level due to the presence of non-planar diagrams. The most
important contribution from non-planar diagrams corresponds to the
$e^+e^-$ pair production
by two virtual photons. The corresponding impact factor contains the
divergency in $s_1$ related to the presence of two-photon
intermediate states in the crossing channel. For the case of $t=0$
it was calculated in ref \cite{KL}). We write down it here only in
the Weizsaekker-Williams approximation, where it has the form of the sum
rule for the Borselino formulae describing the energy dependence of
the total cross-section for the $e^+e^-$ pair production in the
electron-photon collisions:
\begin{eqnarray}
\tau = k^2 \int _{th} ^{s} \frac{d \,s_1}{\pi}\frac{\sigma (s_1)}{s_1}=
a\,ln^2(s/m^2)+b\,ln (s/m^2) +c\,.
\end{eqnarray}
As it was discussed above, the logarithmic dependence on the upper
limit in the integral over $s_1$ should be subtructed in a
self-consistent way to avoid the double counting, because the
logarithmic contributions are summed by the Bethe-Salpeter equation
for the Pomeron in QED. It results
\begin{eqnarray}
\tau(k,0) = \frac{\alpha^3 k^2}{\pi m^2}\left(\frac{418}{27}
-\frac{13}{2}\zeta(2)\right),
\end{eqnarray}
for electron closed loop and
\begin{eqnarray}
\tau(k,0) = \frac{\alpha^3 k^2}{\pi M^2}\left(\frac{3011}{324}
-\frac{28}{9}\zeta(2) - \frac{107}{9}\ln\left(\frac{M}{m}\right)\right),
\end{eqnarray}
for muon closed loop. Here $m,M$ are masses of electron and muon.

For the case of the diagrams without closed fermionic loops
the impact factor appears from the interference of the Bethe-Heitler
amplitudes of the pair production
by virtual photon on electron due to the identity of electrons in
the final state.
It corresponds to the 3-fermion intermediate state in the
$u$-channel of the Compton
scattering amplitude. It was calculated in \cite{KLS}:
\begin{eqnarray}
\tau^{(l)}_{(k,0)}\approx\frac{k^2}{m^2}\frac{\alpha^3}{\pi}
[\frac{221}{315} + \frac{41549}{6300}\xi(2) - \frac{216}{105}\xi(3) -
\frac{792}{105}\xi(2)ln2] \nonumber \\
\approx\frac{k^2}{m^2}\frac{\alpha^3}{\pi}(-3.57).
\end{eqnarray}

The work of EAK was partially supported by RFBR No.~99-02-17730
and HLP No.~99-03.
One of us is grateful to A.E. Dorokhov for help. We also grateful
to M.V. Galynski for collaboration in the initial stage of this paper.

\end{document}